\documentclass[journal=ancac3,manuscript=article,layout=twocolumn]{achemso}

\usepackage{amsmath}
\usepackage{amssymb}
\usepackage{lipsum}
\usepackage{siunitx}
\usepackage[]{todonotes}
\usepackage{svg}
\usepackage{comment}
\usepackage{multirow}
\usepackage{braket}
\usepackage{subfigure}
\usepackage{subcaption}
\usepackage{soul} % for the command \hl
\usepackage{wasysym} % for the diameter command 
\usepackage{tabularx}

\graphicspath{{figures}}

\title{Helium atom micro-diffraction as a characterisation tool for 2D materials}

\author{\underline{Nick A. von Jeinsen}}
\affiliation[Cambridge]
{Cavendish Laboratory, Department of Physics, University of Cambridge, JJ Thomson Ave, Cambridge, UK}
\alsoaffiliation[Ionoptika]
{Ionoptika Ltd., Units B5-B6m Millbrook Close, Chandlers Ford, UK}

\author{\underline{Aleksandar Radi\'{c}}}
\email{ar2071@cam.ac.uk}

\author{Ke Wang}
\author{Chenyang Zhao}
\author{Vivian Perez}
\author{Yiru Zhu}
\author{Manish Chhowalla}
\author{Andrew P. Jardine}
\author{David J. Ward}
\alsoaffiliation[Ionoptika]
{Ionoptika Ltd., Units B5-B6m Millbrook Close, Chandlers Ford, UK}

\author{Sam M. Lambrick}
\email{sml59@cam.ac.uk}
\affiliation[Cambridge]
{Cavendish Laboratory, Department of Physics, University of Cambridge, JJ Thomson Ave, Cambridge, UK}
\alsoaffiliation[Ionoptika]
{Ionoptika Ltd., Units B5-B6m Millbrook Close, Chandlers Ford, UK}


\begin{document}

\begin{abstract}
We present helium atom micro-diffraction as an ideal technique for characterization of 2D materials due to its ultimate surface sensitivity combined with sub-micron spatial resolution. Thermal energy neutral helium scatters from the valence electron density, $\SI{2}-\SI{3}{\angstrom}$ above the ionic cores of a surface, making the technique ideal for studying 2D materials, where other approaches can struggle due to small interaction cross-sections with few-layer samples. Sub-micron spatial resolution is key development in neutral atom scattering to allow measurements from device-scale samples. We present measurements of monolayer-substrate interactions, thermal expansion coefficients, the electron-phonon coupling constant and vacancy-type defect density on monolayer-MoS\textsubscript{2}. We also discuss extensions to the presented methods which can be immediately implemented on existing instruments to perform spatial mapping of these material properties.

\end{abstract}

2D materials represent a large body of research spanning fundamental physics and physical chemistry, to applications in devices including photovoltaics, batteries and transistors. Contained within the 2D material family are numerous classes of materials whose thermal, mechanical and optoelectronic properties vary over orders of magnitude and hence find applicability across the device landscape.
\begin{comment} 
All 2D materials however share that their thickness is only a few Angstroms, resulting in a common difficulty where established non-contact characterisation techniques struggle to measure them due to a lack of interaction cross-section with the probe. Even with relatively low energy probe particles, such as visible photons or low energy electron techniques, 2D materials are readily penetrated, which therefore requires that measurements be corrected for direct substrate interactions with the probe particles. Common techniques that are truly sensitive to only the outermost atoms of a surface, are primarily contact techniques, like STM or AFM. 
\end{comment}
All 2D materials, however, share the characteristic of having a thickness of only a few angstroms. As a result, established non-contact characterization techniques struggle to measure them due to a limited interaction cross-section with the probe. Even when using relatively low-energy probe particles, such as visible photons or low-energy electrons, 2D materials are easily penetrated. Consequently, measurements must be corrected for interactions between the probe particles and the substrate. Techniques that are genuinely sensitive to only the outermost atoms of a surface are primarily contact methods, such as STM or AFM.

An alternative technique is helium atom micro-diffraction presented by von Jeinsen et al. \cite{von_jeinsen_2d_2023}, which uses a thermal energy beam of neutral \textsuperscript{4}He atoms. It has an ultra-low incident energy ($\sim\SI{64}{\milli\electronvolt}$) at ambient temperature, giving the \textsuperscript{4}He a de Broglie wavelength of $\SI{0.06}{\nano\metre}$ commensurate to atomic features, giving the technique ultimate sensitivity. The low beam energy means that the probe particles scatter from the outermost electron density, with a turning point $2-\SI{3}{\angstrom}$ above the ionic cores of the top layer atoms. Therefore, the atoms cannot penetrate the sample and interact directly with the bulk or substrate beneath the true surface atoms. Neutral \textsuperscript{4}He is further advantageous as a probe because it is chemically inert and electrically neutral, making the technique entirely agnostic to sample chemistry, allowing for the measurement of a wide range of sensitive samples without need for coatings or specific sample preparation. With reported spatial resolution reaching $\approx \SI{300}{\nano\metre}$ \cite{nick_mphil_2021}, helium atom micro-diffraction can also measure device scale samples.

In this paper, helium atom micro-diffraction is performed with a Scanning Helium Microscope (SHeM),
\begin{comment} as an emerging technique ideally suited to lab-based 2D material characterization.
\end{comment}
where we present SHeM's current imaging capabilities, in real and reciprocal space, enabling several characterisation methods of 2D materials including, but not limited to, spatial mapping of lattice parameters, contamination, monolayer-substrate interactions, the Debye-Waller factor, vacancy-type defect density and crystal phases. 

\section*{Results and discussion}
\subsection*{Real-space imaging}
\label{sec:real_space_imaging}

\begin{figure*}[t]
    \centering
    \includegraphics[width=0.8
    \linewidth]{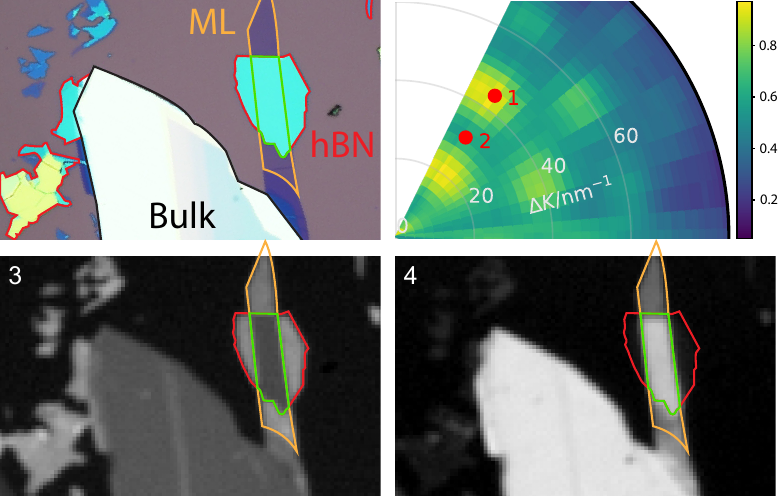}
    \caption{Reflection mode optical (top-left) and real-space SHeM (bottom) images of bulk MoS\textsubscript{2}, hBN/SiO\textsubscript{2} and monolayer MoS\textsubscript{2}/hBN/SiO\textsubscript{2} (Shaded intersection). A 2D micro diffraction measurement (top-right) was taken to identify diffraction conditions for the real space imaging. An obvious change in contrast can be seen as the instrument is configured for }
    \label{fig:real_space_imaging}
\end{figure*}

SHeM's most basic analysis mode rasters the sample laterally in front of the beam to acquire real-space images, forming the basis of all advanced analysis discussed in coming sections. Figure \ref{fig:real_space_imaging} displays real-space images of bulk and monolayer MoS\textsubscript{2}/hBN/SiO\textsubscript{2}, hBN/SiO\textsubscript{2} correlated with reflection mode optical microscopy for reference. Real-spacing imaging is predominantly used for correlation with complementary microscopy techniques to target diffraction measurements, and to investigate topography of macroscopically non-trivial sample geometries \cite{radic_3d_2024,radic_implementing_2021}. It is important to understand that each real-space image, rastered in $(x,y)$, is taken at a single reciprocal-space, or $\Delta K$, value, corresponding to a z-axis position in real-space. Therefore, for clean, single crystalline samples, real-space images will exhibit contrast which is diffractive and topographic simultaneously. By extension, the contribution of diffractive contrast to real-space images means that differing chemical structures or domain orientations can be qualitatively observed immediately from a real-space image.

\subsection*{Surface contamination}

\label{sec:contamination}

\begin{comment}
Helium atom micro-diffraction can be used to detect the presence of surface contaminants using both real-space imaging and diffraction measurements. With a de Broglie wavelength of $\SI{0.06}{\nano\metre}$ at ambient temperatures, the technique is highly sensitive to atomic scale features, which critically includes adsorbates. With the chemically inert beam, and low kinetic energy of (\SI{64}{meV}), the probe neither induces reactions with, nor transfers sufficient momentum to, adsorbates to remove them from the surface. As such, the presented surface analysis measurement of the sample is entirely decoupled from any cleaning procedures one may use, contrary to other non-contact techniques which typically use probe particles of orders of magnitude more energy (e.g. photons or electrons). Contact techniques are also known to be able to desorb or displace surface species due to high scanning tip electric potentials or direct contact with contaminant species (e.g. STM and AFM).
\end{comment}

Helium atom micro-diffraction can be used to detect the presence of surface contaminants using both real-space imaging and diffraction measurements. Importantly, the properties of layered devices are adversely affected by intra-layer contamination. Consequently, it is vital to be able to measure surface cleanliness, contamination, and purity directly on the specific samples intended for device construction. With a de Broglie wavelength of $\SI{0.06}{\nano\metre}$ at ambient temperatures, the technique is highly sensitive to atomic scale features, which critically includes adsorbates. With the chemically inert beam, and low kinetic energy of (\SI{64}{\milli\electronvolt}), the probe neither induces reactions with, nor transfers sufficient momentum to, adsorbates to remove them from the surface. As such, the presented surface analysis measurement of the sample is entirely decoupled from any cleaning procedures one may use, contrary to other more conventional non-contact techniques which typically use probe particles of orders of magnitude more energy (e.g. photons or electrons). Contact techniques are also known to be able to desorb or displace surface species due to high scanning tip electric potentials or direct contact with contaminant species (e.g. STM and AFM) leading to insensitivity, making Helium atom micro-diffraction the best option.

\begin{figure}[h]
    \centering
    \includegraphics[width=\linewidth]{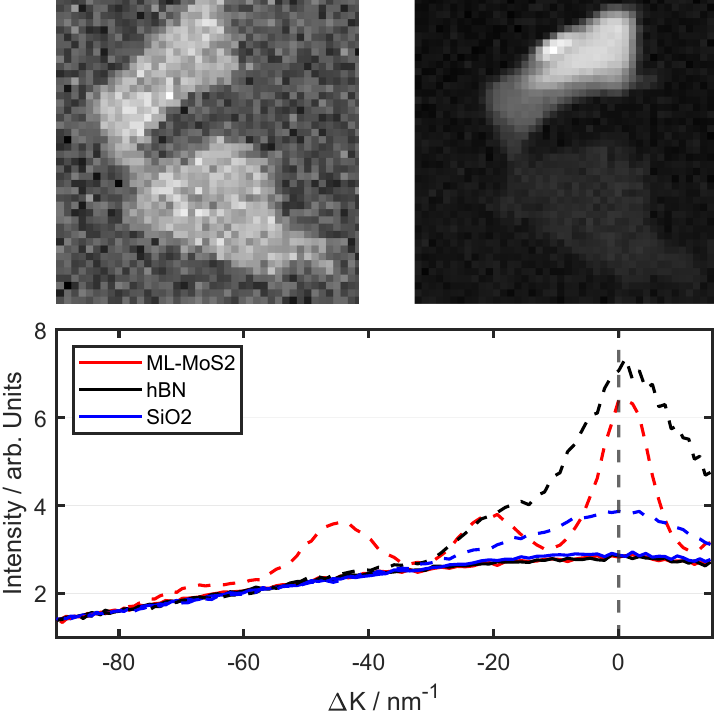}
    \caption{Real-space images and diffraction scans of monolayer-MoS\textsubscript{2}/hBN/SiO\textsubscript{2} (red), few-layer hBN/SiO\textsubscript{2} (black) and SiO\textsubscript{2} (blue) before and after heating (left to right) to $\SI{240}{\degreeCelsius}$ for 8 hours to remove physisorbed species. For monolayer-MoS\textsubscript{2} and hBN we see ordered scattering emerging upon cleaning in contrast to the cosine-like scattering prior. SiO\textsubscript{2} remains cosine-like post-heating. Diffraction scans were acquired at the indicated points in Panels a,b. Real-space images were taken at a single $\Delta K$, marked with a vertical line in Panel C.}
    \label{fig:contaminants}
\end{figure}
The sensitivity of helium scattering to adsorbates has been demonstrated in figure \ref{fig:contaminants} by preparation of a sample containing monolayer-MoS\textsubscript{2} on few-layer hBN on SiO\textsubscript{2}, few-layer hBN on SiO\textsubscript{2} and exposed SiO\textsubscript{2} to provide comparison between two highly ordered, but different, and one amorphous structure when pre- and post-cleaning. The sample was prepared in a glovebox under argon atmosphere and transferred into the SHeM under nitrogen, thus representing typical adsorbate coverage and species that one expects during standard device fabrication. Once under high-vacuum ($\SI{2e-8}{\milli\bar}$) in the SHeM, a real-space image is taken of the as-prepared sample (Figure \ref{fig:contaminants}a) which demonstrates it is difficult to differentiate between MoS\textsubscript{2} and hBN, with the surrounding SiO\textsubscript{2} appearing less intense. Upon heating to $\SI{240}{\degreeCelsius}$ for 8 hours, another real-space image (figure \ref{fig:contaminants}b) reveals a stark difference in contrast between MoS\textsubscript{2}, hBN, and SiO\textsubscript{2}. The MoS\textsubscript{2} and hBN display significantly more intense scattering than the surrounding SiO\textsubscript{2} which can be quantitatively compared in Figure \ref{fig:contaminants}c.

\begin{comment}
Note that the cleaning protocol used is more than the necessary minimum to observe ordered diffraction, we find that most physisorbed species are removed by heating to $\SI{120}{\degreeCelsius}$ for a few hours, but ensures that the surfaces are as clean as possible.
\end{comment}

Having demonstrated the sensitivity of helium atom micro-diffraction to adsorbates, one can trivially extend the method to quantitatively map adsorbate densities across real-space images. By cleaning and measuring the sample with the presented procedure, one can identify the positions of diffraction peaks in reciprocal space ($\Delta K$). This is followed by acquiring a real-space image at a diffraction peak's maximum scattering condition in reciprocal space, achieved by translation of the sample in the z-axis. One can now dose the sample environment with a given contaminant and measure a real-space image to determine if there is a change adsorbate coverage. Through dosing, adsorbation behaviour of a given sample to varying species can also be investigated by monitoring a specific diffraction peak while dosing.

\subsection*{Monolayer-substrate interactions}
\label{sec:substrate}

It is well documented that the choice of substrate placed under a few-layer material has a significant effect on a range of material properties. One can measure the effect of the substrate on optoelectronic or structural properties of the sample \textit{via} methods such as photoluminescence spectroscopy (PL), Raman, or low-energy electron microscopy/diffraction (LEEM/D), respectively\cite{man_protecting_2016}.

\begin{figure}[h]
    \centering
    \includegraphics[width=0.9\linewidth]{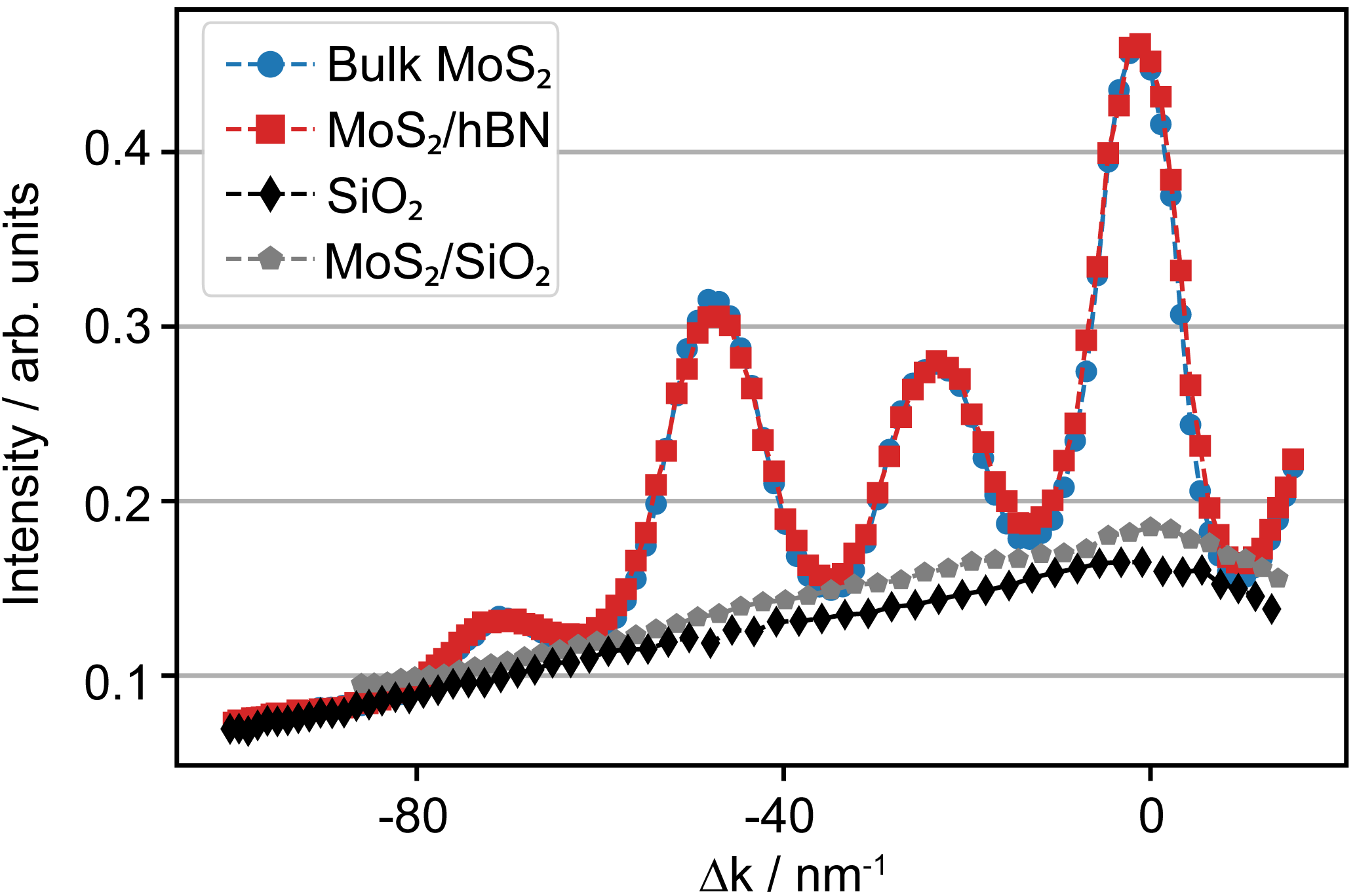}
    \caption{1D diffraction scans showing that the difference in monolayer-MoS\textsubscript{2} structure when placed directly onto SiO\textsubscript{2} (gray) versus when a few-layer hBN buffer is used between the monolayer and SiO\textsubscript{2} (red). Bulk MoS\textsubscript{2} and direct measurement of SiO\textsubscript{2} are included as references for known ordered/disordered scattering, respectively. Figure reproduced with permission from Radi\'{c} et al.\cite{Defects_shem}}
    \label{fig:substrate}
\end{figure}

To demonstrate the exclusive surface sensitivity of helium atom micro-diffraction, monolayer-MoS\textsubscript{2} was placed directly onto both an SiO\textsubscript{2} substrate and few-layer hBN which is in turn on SiO\textsubscript{2}, sample geometry is shown in figure \ref{fig:real_space_imaging}. It has been demonstrated that certain atomically thin buffer layers, such as hBN, LaAlO\textsubscript{3} and SrTiO\textsubscript{3}, can protect the optoelectronic properties of monolayers mounted on them from strongly interacting substrates like SiO\textsubscript{2}\cite{man_protecting_2016}. Figure \ref{fig:substrate} shows that the helium scattering from monolayer MoS\textsubscript{2} becomes almost entirely disordered when mounted directly on SiO\textsubscript{2} (gray), with faint signs of structure remaining at the expected $\Delta K$ diffraction peak positions. In contrast, placing the monolayer onto a few-layer hBN buffer protects its structure and produces diffraction that matches bulk MoS\textsubscript{2} to within experimental bounds, suggesting that their surface morphologies are the same.

We have demonstrated that helium atom micro-diffraction can be used to investigate the effect of substrate choice by direct measurement of monolayer surface morphology. One can leverage the spatial resolution and exclusive surface sensitivity of the technique to investigate inter-layer coupling strength in otherwise difficult to measure systems such as van Der Waals heterostructures where comparable techniques like LEED/M, PL or Raman struggle due to the inherent transmission of the probe through the sample.

\subsection*{Thermal expansion coefficient}
\label{subsec:structure}

Structural information, such as the lattice constant, can be accurately measured using 2D diffraction scans, found to be within $1\%$\cite{von_jeinsen_2d_2023}. 2D diffraction scans offer increased accuracy over 1D scans, like those in figure \ref{fig:contaminants}, because more diffraction provide a statistical advantage.

\begin{comment}
A simple, but key, measurement that can be achieved using diffraction is the surface lattice constant as the diffraction pattern gives the reciprocal lattice. With helium micro diffraction, 2D patterns are best placed to measure the lattice constant due to the large number of diffraction peaks that can be identified. Once a 2D diffraction pattern is acquired the background and peak fitting procedures discussed in section \ref{subsec:analysis_process} can be used to identify the centres of the diffraction patterns -- take for example the diffraction pattern from bulk MoS\textsubscript{2} presented in figure \ref{fig:monolayer_mos2_lattice}, with the identified diffraction peak centres plotted.

Once the diffraction peak centres are identified the reciprocal lattice constant, $G$, can be calculated by averaging all the spacings between the peaks, thus the real space surface lattice constant is found. 
\end{comment}

\begin{figure}
    \centering
    \includegraphics[width=\linewidth]{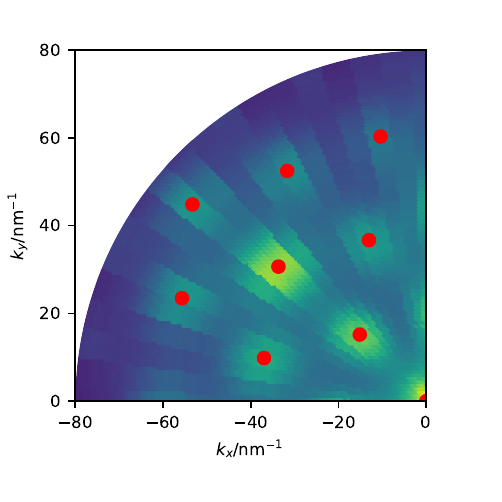}
    \caption{A 2D diffraction pattern taken on monolayer-MoS\textsubscript{2}/hBN/SiO\textsubscript{2}, with the identified centres of the diffraction peaks shown as red dots. All of the lattice spacings between adjacent peaks were measured and averaged give a lattice constant $3.14\pm0.07\si{\angstrom}$.}
    \label{fig:monolayer_mos2_lattice}
\end{figure}

Initially, we measure the 2D diffraction scan on the surface of bulk MoS\textsubscript{2}, shown in figure S1, and determine the lattice constant to be $3.15\pm0.07\si{\angstrom}$, matching literature values within experimental uncertainty. We extend the method to mechanically exfoliated monolayer-MoS\textsubscript{2}/hBN/SiO\textsubscript{2}, with diffraction pattern shown in figure \ref{fig:monolayer_mos2_lattice}, to showcase the exclusive surface sensitivity of the technique and determine that the lattice constant is $3.14\pm0.07\si{\angstrom}$, which is within experimental error bounds of accepted literature values.

The thermal expansion coefficient (TEC) of 2D materials is crucial to understanding their behaviour for successful integration into optoelectronic devices, these materials also provide direct measurement of the anharmonic vibrational modes of low-dimensional systems\cite{Late2014,Su2014}, however, difficult to measure due to their optical transparency\cite{Geim2007,Zhang2019}. Using helium atom micro-diffraction one can measure the lattice constant, and therefore TEC, of microscopic optically transparent materials, whether supported by a substrate or free-standing. 

The diffraction pattern was measured along the principle $\braket{10}$ azimuth of MoS\textsubscript{2} at temperatures from $\SI{60}-\SI{450}{\degreeCelsius}$ (the data is presented in figure \ref{fig:monolayer_mos2_lattice}), as with work on HAS no change in the lattice constant was observed, allowing us to put an upper bound on the expansion coefficient of $\SI{16e-6}{\per\kelvin}$. Our measurement is consistent with the upper bound determined as $\alpha<\SI{14e-6}{\per\kelvin}$ reported by Anemone et al.\cite{anemone_setting_2022}, and $\alpha=\SI{7.6e-6}{\per\kelvin}$ measured by Zhang et al. using micro-Raman spectroscopy\cite{Zhang2019}. 

Our method is currently limited in angular resolution (radial dimension of figure \ref{fig:monolayer_mos2_lattice}) by the size of the limiting aperture that lies between the sample and detector. The current configuration uses a nominal $\SI{0.5}{\milli\metre}$ diameter hole for an in-plane angular resolution $\sim\SI{7.9}{\degree}$, representing a compromise between measurement time and helium signal. By changing the limiting aperture to a rectangle with dimensions $0.075\times\SI{0.5}{\milli\metre}$ we can integrate the signal in the axis perpendicular to any change in lattice parameter, the radial direction in figure \ref{fig:monolayer_mos2_lattice}, while improving in-plane angular resolution. We estimate that moving from the current circular aperture to the proposed rectangular aperture decreases helium signal by a factor of $5$, but improves in-plane angular resolution by a factor of $6.7$ from $\SI{7.9}{\degree}\rightarrow\SI{1.2}{\degree}$ which would be sufficient to resolve the TEC over the current temperature range. Details of the current instrument's spatial and angular resolutions, alongside their definitions, can be found in section `Experimental Methods - Helium atom micro-diffraction instrument details'.

\begin{comment}
    Upper limits have been put on the coefficient of thermal expansion for MoS\textsubscript{2}, PtTe\textsubscript{2} and Gr/Cu(111)\cite{anemone_setting_2022,anemone_experimental_2018}.
\end{comment}

\subsection*{Electron-phonon coupling}
\label{sec:DW_factor}

Electron-phonon coupling is a major decoherence mechanism in semiconductors, often causing electron-phonon scattering, and eventually energy dissipation, in turn harming optoelectronic device performance. Optimisation of electron-phonon coupling is therefore critical in device design, both in cases where it must be minimised, or maximised. Electron-phonon coupling is typically measured using inelastic optical techniques, such as Raman or Brillouin scattering, and are primarily limited to optically transparent materials and access to electronic coupling to optical phonons only. It has been shown that helium atom scattering is an effective tool for determining the coupling strength between 2D materials and their substrates \textit{via} Debye-Waller attenuation of helium scattering \cite{AlTaleb2018,anemone_experimental_2018,anemone_electronphonon_2019,anemone_electronphonon_2021}. However, previous atom scattering techniques have been limited to millimetre-scale spot sizes which made device-scale samples inaccessible. Here we introduce helium atom micro-diffraction as a technique well suited to the measurement of electron-phonon coupling in a range of materials where its non-damaging, surface sensitive and microscopic neutral probe can be used to access electronic couplings to both optical and acoustic phonon modes in 2D materials. 

\begin{comment}
Using the sample heater shown in Section \ref{subsec:heater} even relatively strongly absorbing atmospheric contaminants, like water, can be desorbed from the sample, leaving a clean surface for diffraction measurements. Alongside investigations into crystallography, surface contaminants and substrate effects detailed in Sections \ref{subsec:lattice_constants}, \ref{subsec:contamination} and \ref{subsec:substrate_effects}, sample heating also enables spatial mapping of Debye-Waller attenuation with sub-micron lateral resolution.
\end{comment}

The temperature dependence of reflected and diffracted helium intensities can be modelled by Debye-Waller attenuation \cite{AlTaleb2018,anemone_experimental_2018}, which describes the increasing motion of the surface atoms with temperature and causes an increase in inelastic scattering. Thus, the ordered scattering intensity is reduced. The attenuation is described by
\begin{equation}\label{eq:D-W}
    I/I_0 = \exp\left(-2TW\right)
\end{equation}
where $2W$ is the D-W factor. As the helium scattering occurs from the valence electron density rather than from the ionic cores themselves, one can link the D-W factor to electron phonon coupling, as described by Al Taleb et al.\cite{anemone_experimental_2018}. D-W attenuation has been measured previously for macro-scale samples for both LiF and bulk MoS\textsubscript{2}, however without the spatial resolution enabled by SHeM, monolayer MoS\textsubscript{2} flakes could not be measured\cite{anemone_experimental_2018}.
\begin{comment}
As with other types of helium diffraction experiments, measurements of the Debye-Waller factor have previously been limited to large single crystal samples, but with the introduction of micro-atom diffraction they can now be performed on sub-micron spot sizes.
The D-W factor from a LiF crystal cleaved in air was used as a test of the procedure using the Cambridge B-SHeM, alongside measurements on bulk and monolayer MoS\textsubscript{2}. D-W attenuation has been measured previously for macro-scale samples for both LiF and bulk MoS\textsubscript{2}, however without the spatial resolution enabled by SHeM monolayer MoS\textsubscript{2} could not be measured.
\end{comment}

\begin{figure}
    \centering
    \includegraphics{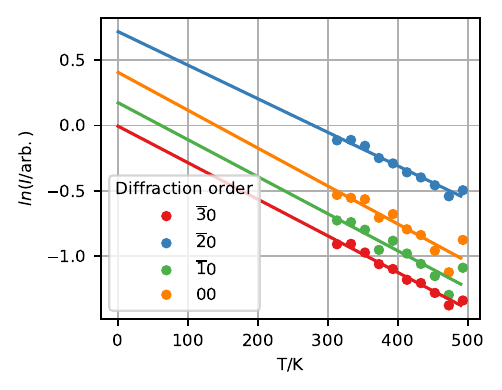}
    \caption{Log diffracted intensity as a function of temperature for monolayer MoS\textsubscript{2} on few-layer hBN on glass. In theory all diffraction orders should yield the same D-W, and therefore electron-phonon coupling, constants but the signal-to-noise ratios of each peak are a function of relative peak heights and instrument geometry, resulting in small differences.}
    \label{fig:dw_factor}
\end{figure}

From figure \ref{fig:dw_factor} we can extract the D-W factor as the straight-line gradient, and from that the electron-phonon coupling constant $\lambda$, contained in table \ref{tab:dw_factor} for monolayer and bulk MoS\textsubscript{2}. The exponents are compared relative to each other, and to literature values for the bulk case, and found to be in agreement with the expected behaviour with parallel momentum transfer ($\Delta K$).    

\begin{table}
\centering
\setlength{\tabcolsep}{2pt} %
\begin{tabular}{c|c|c|c}
    \hline
    MoS\textsubscript{2} & Diffr. Order & D-W /$\SI{e-3}{\per\kelvin}$ & $\lambda$/$\pm\SI{15}{\percent}$\\\hline
    \multirow{4}{*}{\centering ML} & $00$ & $-2.9\pm0.5$ & 0.41\\
    & $\overline{1}0$ & $-2.8\pm0.4$ & 0.40\\
    & $\overline{2}0$ & $-2.6\pm0.2$ & 0.37\\
    & $\overline{3}0$ & $-2.8\pm0.2$ & 0.40\\\hline
    \multirow{4}{*}{\centering Bulk} & $00$ &  $-2.8\pm0.2$ & 0.51\\
    & $\overline{1}0$ & $-2.6\pm0.2$ & 0.47\\
    & $\overline{2}0$ & $-2.8\pm0.4$ & 0.51\\
    & $\overline{3}0$ & $-2.9\pm0.4$ & 0.53\\
    \hline
\end{tabular}
\caption{Debye-Waller factors and electron-phonon coupling constants extracted from the data presented in figure \ref{fig:dw_factor} using the model in equation \ref{eq:D-W} and theoretical relations outlined by Anemone et al.\cite{anemone_experimental_2018}, respectively.}
\label{tab:dw_factor}
\end{table}

We report the electron-phonon coupling constant in bulk MoS\textsubscript{2} as $\lambda_{bulk}\approx\SI{0.51e-3}{\per\kelvin}$, showing good agreement to the literature values $\lambda\approx\SI{0.41e-3}{\per\kelvin},\SI{0.49e-3}{\per\kelvin}$ measured using a typical helium atom scattering instrument with few-millimetre spot size, reported by Anemone et al.\cite{anemone_experimental_2018}. We also find $\lambda_{ML}\approx\SI{0.40e-3}{\per\kelvin}$ for ML-MoS\textsubscript{2} on few-layer hBN. We find the electron-phonon coupling constant is $\sim\SI{20}{\percent}$ smaller in the monolayer compared to bulk MoS\textsubscript{2}. By leveraging the current best reported $\approx \SI{350}{\nano\metre}$ spot size, the method can be extended to perform spatial mapping of the Debye-Waller factor, and therefore electron-phonon coupling constant, without sample preparation or damage in thin films and delicate materials, a class of materials typically difficult to characterise using standard optical and electron beam techniques.

\subsection*{Vacancy-type defect density}
\label{sec:defects}
\begin{comment}
The work in this section was presented in detail by Radic et al. ??? \cite{??}.
Defects in 2D materials are of great interest in their ability to functionalize materials, in particular they can be used to tune the optoelectronic properties of TMDs[??]. However introducing too many defects results in either a degradation of the material itsle or a reduction in the electrical conductance of the material[??]. Therefore carefully tuning the defect density in TMDs is an outstanding problem, with no lab based instrument capable of determining defect density[??]. Recently a new, simple, method of inducing defects in MoS\textsubscript{2} has been developed which allows the straightforward preparation of Monolayer MoS\textsubscript{2} with different densities of sulphur vacancies[??]. Helium scattering is known to have a large cross section to atomic scale surface defects and therefore atom micro-diffraction is a promising tool for 
\end{comment}

Precise control of defect density in semiconductors is instrumental for both current and future semiconductor device  development. In particular, the optoelectronic properties of two-dimensional TMD semiconductors such as MoS\textsubscript{2}, can be tuned using single-atom defects\cite{Regan2022}. Applications of these materials includes catalysis\cite{yang_single_2019} and a plethora of devices\cite{mitterreiter_role_2021,Barthelmi2020,Chakraborty2019,LopezSanchez2013}, all of which can affected by the material's defect density.
In many applications a balance must be made between a sufficiently high number of defects for the material to acquire the desired properties while the material remaining sufficiently ordered to not degrade electronic performance\cite{zhu_room-temperature_2023}, or even entirely degrade the lattice structure. However, quantification of defect densities in 2D materials remains a significant experimental challenge, where typically used methods are XPS \cite{zhu_room-temperature_2023} and STEM, with conductive AFM (CAFM) being explored recently\cite{defects_mos2_cafm}. As all of these methods commonly require complicated sample preparation processes, there is a characterisation shortcoming that is only going to grow more acute as devices using 2D materials start being produced on an industrial scale, and therefore enter the commercial sphere. In this work we demonstrate how helium atom micro-diffraction can be used to characterise the vacancy-type defect density on the surface of few-layer materials using data reproduced from Radi\'{c} et al. with permission\cite{Defects_shem}.

\begin{comment}
Defects in a crystalline surface cause a local region of disorder in the otherwise ordered surface, and thus contribute to a reduction in the intensity of helium flux that is scattered into Bragg diffraction channels. Instead, the defect scatters the atom flux randomly, contributing to a diffuse signal with a cosine distribution centred on the normal to the surface, analogous to Lambertian scattering in optics\cite{lambrick_observation_2022,Lambert1760}. Therefore an increasing defect density, $\Theta$, will result in a decreased atom diffraction intensity, The relationship is quantified \cite{??} according to the defect cross-section, sigma:
\begin{equation}
I/I_0 = (1 – \Theta)^{\sigma c},
\label{eqn:defect_scattering_cross_section}
\end{equation}
Whereby reference to the diffracted intensity from a sample with intrinsic defect density, $I_0$, the defect density in a new sample can be inferred from $I$. Equation \ref{eqn:defect_scattering_cross_section} holds for defect densities which are sufficiently low that defects can be considered as isolated from one another. Once defect density becomes sufficiently high that scattering cross-sections of individual defects begin to overlap, the number of bi-vacancies becomes non-negligible, and the model breaks down.
\end{comment}

Three mechanically exfoliated monolayer flakes of MoS\textsubscript{2} with increasing defect densities, ranging from the intrinsic $\SI{0.1e14}{\per\centi\metre\squared}$ to $\SI{1.8e14}{\per\centi\metre\squared}$, and $\approx\SI{15}{\micro\metre}$ lateral size were produced using high-temperature annealing under a mixed Ar/H\textsubscript{2} ($95\%/5\%$) atmosphere as outlined by Zhu et al.\cite{zhu_room-temperature_2023}. Exact annealing parameters, and the resulting defect densities inferred using photoluminescence spectroscopy, which has been calibrated against stoichiometric beam-line XPS, are contained in Table \ref{tab:annealing_protocol}.
\begin{figure}[h]
	\centering
	\includegraphics[width=\linewidth]{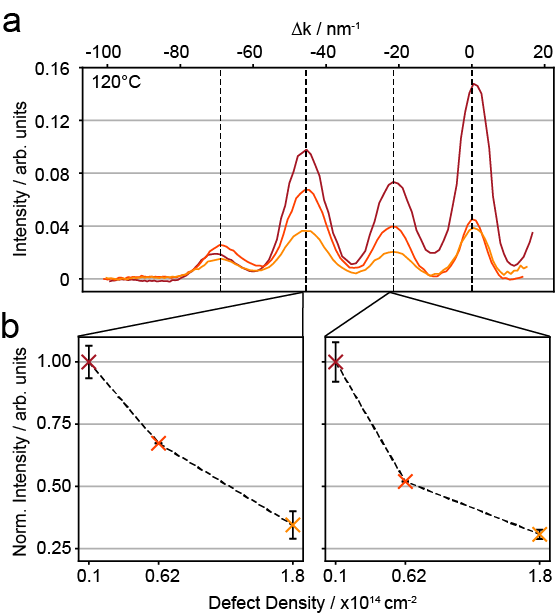}
	\caption{Helium atom micro-diffraction scans along a principle $\braket{10}$ azimuth of mechanically exfoliated monolayer MoS\textsubscript{2} flakes. Taking line cuts through the 2\textsuperscript{nd} and 3\textsuperscript{rd} order diffraction peaks shows that increasing defect density approximately linearly decreases diffracted intensity. Figure reproduced with permission from Radi\'{c} et al.\cite{Defects_shem}}
	\label{fig:defect_measurement}
\end{figure}

\begin{comment}
    
Figure \ref{fig:defect_measurement} shows that the diffracted intensity decreases approximately linearly with increasing point-defect density in monolayer MoS\textsubscript{2} as predicted by
\begin{equation}
I/I_0 = (1 – \Theta)^{\sigma n},
\label{eqn:defect_scattering_cross_section}
\end{equation}

where $\Theta$ is the defect density expressed as a fraction of the available sites on the surface, $\sigma$ is the \textsuperscript{4}He-defect scattering cross-section and $n$ is the unit cell area.
\end{comment}

Figure \ref{fig:defect_measurement} shows that diffracted helium intensity decreases approximately linearly as a function of defect density within the single-vacancy limit. Detailed analysis and discussion of the results in figure \ref{fig:defect_measurement} can be found in the full manuscript\cite{Defects_shem}.

The presented work on monolayer MoS\textsubscript{2} validates that microscopic helium atom diffraction can be employed as a lab-based method to quantify point-defect density without specific sample preparation or damage. The method is agnostic to sample chemistry, and thickness, because the mechanism through which it measures defect density is purely geometric, relying upon the degree of order of the sample surface, meaning that the method can be trivially extended to any system whose macroscopic properties are mediated by surface defects or dopants. Examples of current systems include hBN\cite{Stern2024}, graphene\cite{Bhatt2022}, doped systems such as diamond \cite{Einaga2022}, alongside other TMDs. In conjunction with its electrostatically neutral and low energy beam, microscopic helium atom diffraction presents itself as an ideal technique for the investigation of these few-layer materials. Further improvements in the instruments' lateral resolution resolution while performing diffraction measurements, down to the previously reported $\sim\SI{300}{\nano\metre}$ spot sizes\cite{nick_mphil_2021}, would allow for sub-micron scale mapping of defect density across samples. The technique is not currently capable of acting as a standalone, stoichimetric measure of defect density, like XPS, once an accurate helium-defect scattering cross-section is calculated this will also be possible.

\section*{Conclusion}
\label{sec:conclusion}

The findings reported here have demonstrated the effectiveness of helium atom micro-diffraction as a powerful, and completely non-invasive characterisation tool for 2D materials. The technique’s ability to probe only the outermost atomic layers with micron spatial resolution makes it uniquely suited to addressing the challenges of analysing 2D materials, where established techniques often struggle.

The results and discussion have highlighted several key applications of helium atom micro-diffraction, including the measurement of surface cleanliness and contamination without altering the sample, a critical feature for ensuring the integrity of materials used in device fabrication. We also investigated the impact of substrate on monolayer properties, confirming that the use of a few-layer hBN buffer can preserve the structural integrity of monolayer MoS2, whereas direct contact with SiO2 results in significant disorder \cite{electron_diffraction_paper}.

We have also explored the potential of helium atom micro-diffraction for determining the thermal expansion coefficient and electron-phonon coupling in monolayer MoS2, and have demonstrated its capability for precise structural and thermal analysis. The technique’s ability to quantify defect densities without sample preparation or damage also highlights it as an obvious tool for optimizing 2D materials in ever expanding areas of application, from optoelectronics to catalysis and more \cite{}.

Future enhancements to helium atom micro-diffraction, such as achieving sub-50 nm spatial resolution, and incorporating out-of-plane scattering capabilities, will further expand its applicability, particularly for complex heterostructures and device-grade materials. In summary,  helium atom micro-diffraction promises to provide a robust and adaptable platform for characterisation and development of 2D materials, paving the way for advancements in nanotechnology and material science.

\section*{Experimental Methods}
\label{sec:methods}

\subsection*{Helium atom micro-diffraction instrument details}
\label{sec:instrument}

In the presented work we used a Scanning Helium Microscope (SHeM) with $\SI{5}{\micro\metre}$ spatial and $\SI{7.9}{\degree}$ in-plane angular resolution at the specular condition, representing a factor of 2 improvement in spatial resolution over the first iteration published by von Jeinsen et al.\cite{von_jeinsen_2d_2023}. Spatial resolution is defined as the full-width at half-maximum of the beam spot on the sample at the designed working distance. In-plane angular resolution is given by the angle subtended by the limiting detector aperture and beam spot on the sample, with the correcting factor $\frac{1}{\sqrt{2}}$ applied to account for the $\SI{45}{\degree}$ instrument geometry.

\subsection*{Sample details}
\label{sec:sample_details}

Monolayers of MoS\textsubscript{2} were produced by mechanical exfoliation of a bulk MoS\textsubscript{2} crystal, purchased from 2D Semiconductors Ltd., and were deposited onto few-layer thick hexgonal boron nitride (hBN) which  in turn lies on a SiO\textsubscript{2} substrate (high-precision glass microscope slide). Each monolayer measured $\approx\SI{15}{\micro\metre}$ laterally.

The increased defect density monolayer MoS\textsubscript{2} samples, presented in section \ref{sec:defects}, were produced the same as previously described in this section with the addition of a thermal annealing process under a mixed Ar/H\textsubscript{2} $(95\%/5\%)$ atmosphere for $\SI{0.5}{\hour}$, following the method as described by Zhu et al. \cite{zhu_room-temperature_2023}. Varying annealing temperature and time allows for control of defect density. Exact annealing parameters, with final defect densities characterised by stoichiometric XPS, are shown in Table \ref{tab:annealing_protocol}.

% \usepackage{multirow}
\begin{table}[]
\setlength{\tabcolsep}{2pt} %
\begin{tabular}{|c|cc|c|}
\hline
\multirow{2}{*}{Sample} &
  \multicolumn{2}{c|}{Annealing Protocol} &
  \multirow{2}{*}{\begin{tabular}[c]{@{}c@{}}Defect Density \\  /$\SI{e14}{\per\centi\metre\squared}$\end{tabular}} \\ \cline{2-3}
   & \multicolumn{1}{c|}{Time / hrs} & Temp. / C &                    \\ \hline
1a & \multicolumn{1}{c|}{N/A}        & N/A       & $\sim$0.1 (native) \\ \hline
2  & \multicolumn{1}{c|}{0.5}        & 550       & 0.62               \\ \hline
3  & \multicolumn{1}{c|}{0.5}        & 600       & 1.8                \\ \hline
\end{tabular}
\caption{Annealing protocols used to generate increased defect density monolayer MoS\textsubscript{2} samples shown in figure \ref{fig:defect_measurement}. Table has been reproduced with permission from Radi\'{c} et al. \cite{Defects_shem}.}
\label{tab:annealing_protocol}
\end{table}

\section*{Acknowledgements}

The work was supported by EPSRC grant EP/R008272/1, Innovate UK/Ionoptika Ltd. through Knowledge Transfer Partnership 10000925. The work was performed in part at CORDE, the Collaborative R\&D Environment established to provide access to physics related facilities at the Cavendish Laboratory, University of Cambridge and EPSRC award EP/T00634X/1. SML acknowledges support from EPSRC grant EP/X525686/1.

\bibliography{micro-diffract}
%\newpage
%\input{supporting_information}

\end{document}



\setcounter{figure}{0}
\renewcommand{\thefigure}{S\arabic{figure}}

\section{Bulk MoS\textsubscript{2} diffraction pattern}

\begin{figure}[h]
    \centering
    \includegraphics[width=\linewidth]{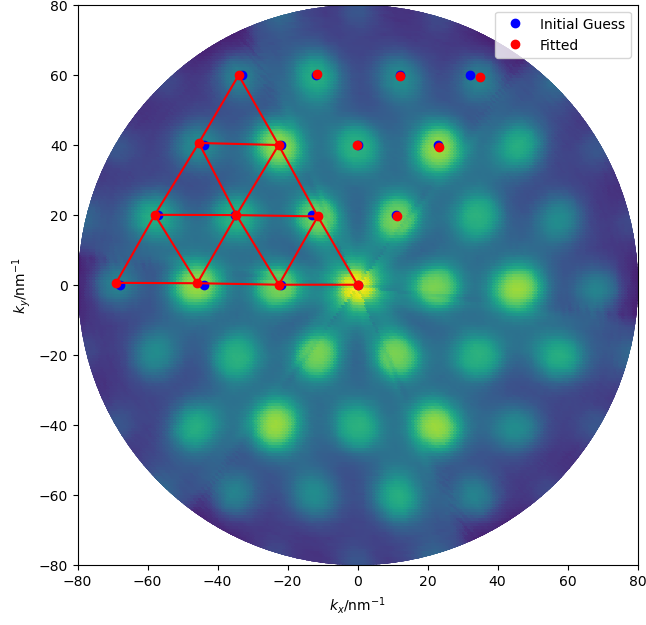}
    \caption{A 2D diffraction pattern taken on the surface of bulk MoS\textsubscript{2} from which the lattice constant is measured as $3.15\pm0.07\si{\angstrom}$.}
    \label{fig:bulk_mos2_lattice}
\end{figure}

\begin{comment}
    
\section{Point Defect Density Monolayer MoS\textsubscript{2} Sample Details}
\label{app:defects_details}

\begin{table}[h]
\begin{tabular}{|l|ll|l|}
\hline
\multicolumn{1}{|c|}{\multirow{2}{*}{Sample ID}} & \multicolumn{2}{c|}{Annealing Protocol} & \multicolumn{1}{c|}{\multirow{2}{*}{\begin{tabular}[c]{@{}c@{}}Defect Density\\ /$\num{1e14}\,cm^{-2}$\end{tabular}}} \\ \cline{2-3}
\multicolumn{1}{|c|}{} & \multicolumn{1}{c|}{Time / hrs} & \multicolumn{1}{c|}{Temp. / C} & \multicolumn{1}{c|}{} \\ \hline
1a & \multicolumn{1}{l|}{N/A} & n/A & $\sim$0.1 (intrinsic) \\ \hline
2 & \multicolumn{1}{l|}{0.5} & 550 & 0.62 \\ \hline
3 & \multicolumn{1}{l|}{0.5} & 600 & 1.8 \\ \hline
\end{tabular}
\caption{Annealing parameters used to induce sulfur-vacancy points defects in MoS\textsubscript{2} following the method outlined by Zhu et al. using a flowing H2/Ar atmosphere under high temperature\cite{zhu_room-temperature_2023}.}
\label{tab:defect_sample_details}
\end{table}
\end{comment}

\section{Temperature dependence of diffracted intensity in monolayer MoS\textsubscript{2}}
\label{app:D-W_mos2}

\begin{figure}[h]
    \centering
    \includegraphics[width=\linewidth]{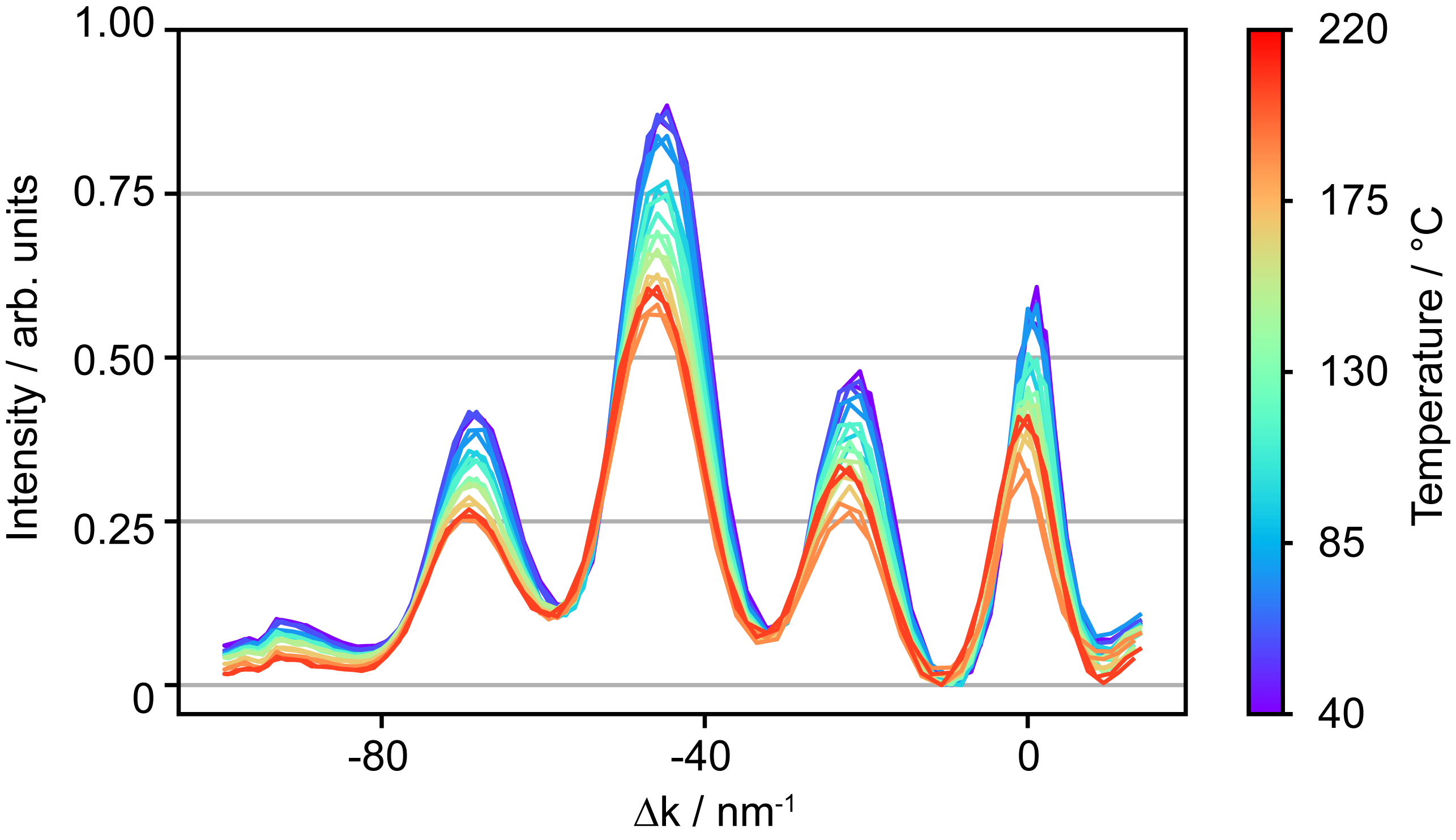}
    \caption{Micro helium atom diffraction scans of an MoS\textsubscript{2} monolayer with intrinsic defect density taken along the principle $\braket{10}$ azimuth as a function of temperature. A clear inverse relationship between diffracted intensity and temperature is visible, the Debye-Waller factor, followed by the electron-phonon couplig constant, can be extracted as shown in figure 5, with exact values for all diffraction peaks in table 1.}
    \label{fig:mos2_temp_dependence}
\end{figure}